\documentclass[10pt,letterpaper]{article}
\usepackage{opex3}
\usepackage{amsfonts}
\usepackage{amsmath}
\usepackage{graphicx}
\usepackage{cite}

\newcommand\OEtitleB[1]{\LARGE \bf \hskip0.0pc \parbox{1.0\textwidth}{ \noindent%
   \LARGE \bf \begin{center} #1 \end{center}\rm } \vskip.1in \rm\normalsize }
   
\let\title\OEtitleB

\begin{document}

\title{Narrow band amplification of light carrying orbital angular momentum}

\author{G. C. Borba${^1}$, S. Barreiro$^{1,3}$ L.~ Pruvost${^2}$, D.~Felinto${^1}$, and J.~W.~R.~ Tabosa${^{1,*}}$}

\address{$^{1}$Departamento de F\'{\i}sica, Universidade Federal de Pernambuco, 50670-901 Cidade Universit‡ria, Recife, PE-Brazil \\
$^2$Laboratoire Aim\'{e}-Cotton, CNRS, Universit\'{e} Paris-Saclay, ENS Cachan, 91405 Orsay, France\\
$^3$Unidade Academica do Cabo de Santo Agostinho, Universidade Federal Rural de Pernambuco, BR 101 SUL, PE-Brazil\\}
\email{$^*$tabosa@df.ufpe.br}

\begin{abstract}
We report on the amplification of an optical vortex beam carrying orbital angular momentum via induced narrow Raman gain in an ensemble of cold cesium atoms. A 20\% single-pass Raman gain of a weak vortex signal field is observed with a spectral width of order of 1 MHz, much smaller than the natural width, demonstrating that the amplification process preserves the phase structure of the vortex beam. The gain is observed in the degenerated two-level system associated with the hyperfine transition $6S_{1/2}(F=3)\leftrightarrow 6P_{3/2}(F^{\prime}=2)$ of cesium.  Our experimental observations are explained with a simple theoretical model based on a three-level $\Lambda$ system interacting coherently with the weak Laguerre-Gauss field and a strong coupling field, including an incoherent pumping rate between the two degenerate ground-states.   
\end{abstract}


\section{Introduction}

\noindent Several applications of light beams carrying orbital angular momentum (OAM) \cite{Allen04, Torner13} have been proposed, including their use to enhance the efficiency of quantum information and computation protocols through the multidimensional state space spanned by OAM modes \cite{Bechmann-Pasquinucci00}. More recently these beams have also been used to demonstrate the enhancement of classical wireless communication in the millimeter-wave range \cite{Willner14}, and to implement high rate data encoding \cite{Willner15}. The modes of light with OAM can provide many discrete states for encoding information. A well-known family of vortex beams is constituted by Laguerre-Gaussian (LG) modes specified by a topological charge $\ell$, which gives to the mode a helicoidal phase structure and a corresponding OAM per photon equal to $\ell\hbar$ \cite{Allen92}.

	Besides these recent demonstrations of classical communication using OAM modes, previous work reporting the storage and manipulation of OAM of light have also been published, both in the classical \cite{Pugatch07, Moretti09} and quantum regimes  \cite{Laurat13}. Also of considerable importance for communication schemes using OAM encoded information is the possibility of retrieving the stored information along different directions in space, as well as the manipulation, through higher-order nonlinearity, of the stored information, which have also been demonstrated  \cite{Rafael14, Rafael15}. Still in the context of communication based on encoded OAM information, another important issue is the possibility of amplification of the encoded signal, which according to our knowledge has not been demonstrated yet. As it is well known, long distance optical communication systems are strongly affected by propagation losses, and amplifying nodes are usually  required to recover the optically encoded information. 
	
	In this article we address this issue by demonstrating that a LG mode can be amplified through a nonlinear stimulated Raman process. The nonlinear interaction of a LG beam with an atomic ensemble has been investigated previously by different groups  \cite{Barreiro03, Guo06} mainly through four-wave mixing (FWM) processes. Although the FWM process can produce parametric gain in a weak signal beam, it usually requires specific experimental conditions of high intensity and far detuned light beams \cite{Marino08}. In the present work we use ground-state coherence between Zeeman sublevels of the cesium hyperfine ground state $6S_{1/2}(F=3)$, induced by  nearly resonant coupling and signal beams to obtain $cw$ gain on the signal beam. Similar gain mechanism has been investigated before by many authors and modeled using  a three-level $\Lambda$ scheme driven by two coherent fields in presence of optical pumping induced by the strong coupling beam, which in this case needs to interact with both transitions of the  $\Lambda$ system \cite{Kumar85, Chiao00}. Previous related works have also investigated gain in three-level $\Lambda$ scheme in the context of laser without inversion ($LWI$), where an extra incoherent \cite{Scully96} or coherent \cite{Zhu96} pumping beam is used. 
	
	It is also worth mentioning that narrow spectral gain in multilevel atomic system, as for example a degenerate two-level system associated with a hyperfine transition $F\rightarrow F+1$, interacting with two red-detuned orthogonally polarized laser fields, has been  first observed and interpreted in terms of Raman transitions between unequally populated Zeeman sublevels in Refs.  \cite{Tabosa91, Grison91}. However, it is well known that for a $F\rightarrow F-1$ transition interacting with two orthogonally polarized laser fields, the coupling and the signal fields, the existence of a dark state only leads to the observation of electromagnetically induced transparency ($EIT$) \cite{Lezama99}. In our present experiment we add an independent laser beam which leads to the creation of a population inversion between the Zeeman ground state sublevels and is responsible for the amplification of the signal field via stimulated Raman transition. Finally, we would like to emphasize that the observed gain mechanism is not associated with parametric FWM, since we have experimentally verified it is insensitive to phase matching, and can be observed without the presence of FWM signal.

\section{Experiment and analysis}

\subsection{Raman gain spectrum}

The experiment is performed in cold cesium atoms obtained from a MOT, using the cesium hyperfine two-level degenerate transition $6S_{1/2}(F=3)\leftrightarrow 6P_{3/2}(F^{\prime}=2)$. The atoms are initially pumped into the ground state $6S_{1/2}(F=3)$  via a non resonant excitation induced by the trapping beams, by switching off  the MOT repumping beam and the quadrupole magnetic field $ 1 ms$ before the switching off of the trapping beams. This allow us to obtain an optical density of order of 3, in an atomic sample of size $\approx$ 2 mm. Three pairs of Helmholtz coils are used to compensate stray magnetic fields. On the cold atomic ensemble a coupling beam $C$ and a signal beam $S$ are incident forming a small angle $\theta\approx 2^o$ and having orthogonal linear polarizations as indicated in Fig. 1(a). As it is also indicated in Fig. 1(a), a nearly couterpropagating pumping beam $P$, having the same linear polarization as the signal beam $S$,  is also incident on the atomic ensemble. All the beams are kept on for about $30 \mu s$, so to allow the system to reach steady state. The frequencies of these three beams can be controlled independently by acousto-optic modulators (AOM). For this polarization configuration of the incident beams one can excite different sets of $\Lambda$ three-level systems with two degenerate Zeeman ground states and one excited state, as shown in  Figs. 1(b), which shows a generic set of Zeeman sublevels interacting with the incident fields, where we are considering the quantization axis along the coupling beam polarization direction. \\

\begin{figure}[htb]
  \centerline{\includegraphics[width=10.5cm]{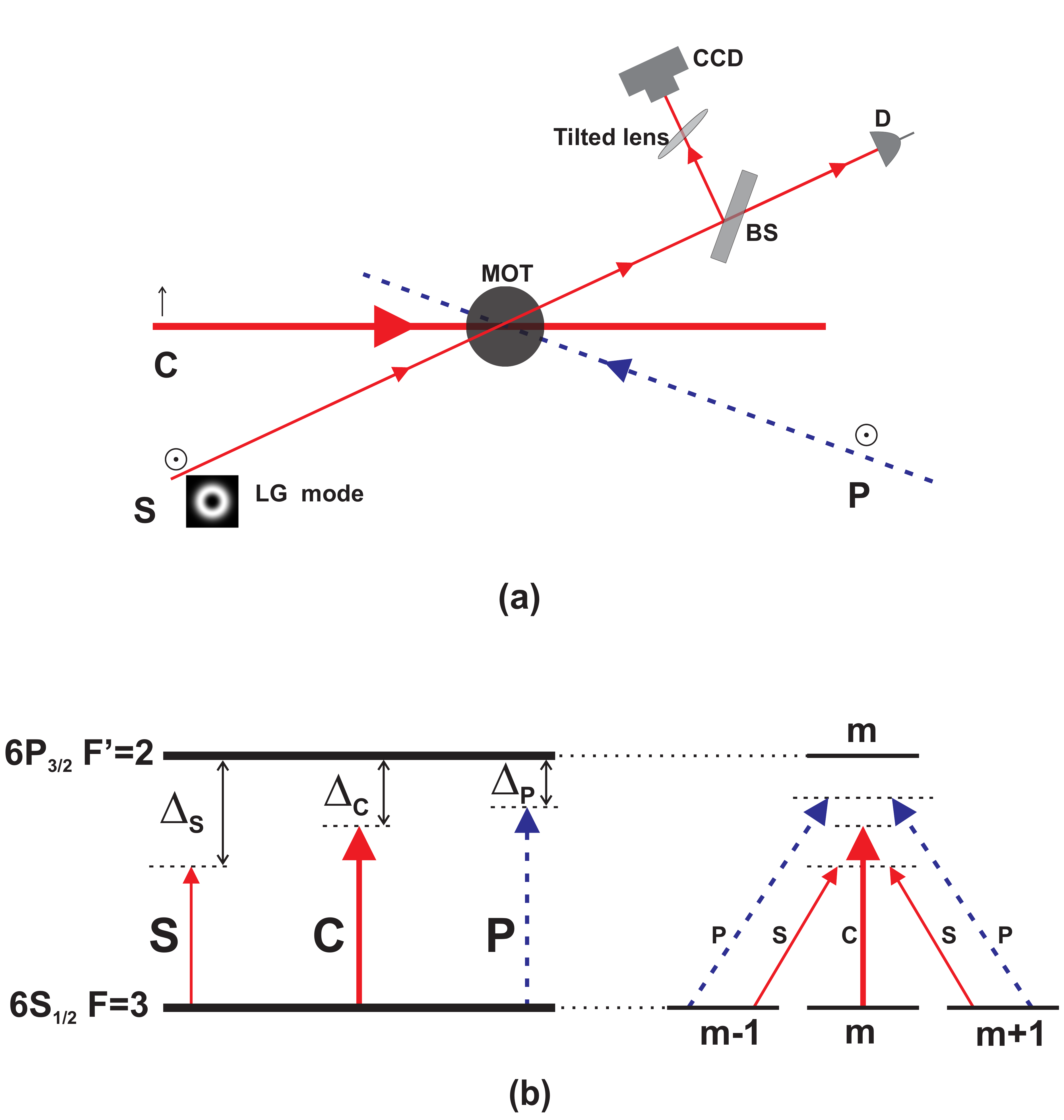}}
  \vspace{0.5cm}
  \caption{(a) Simplified experimental scheme showing the propagation directions of the incident beams ($C$, $S$ and $P$). The direction of beam $P$ is chosen to eliminate the FWM parametric gain. (b) The degenerate two-level system of cesium, corresponding to the hyperfine transition $(F=3)\leftrightarrow (F^{\prime}=2)$, showing a generic set of Zeeman sublevels and indicating the coupling of the sublevels with the respective optical fields.}
\label{fig:Fig1}
\end{figure}

\begin{figure}[htb]
  \centerline{\includegraphics[width=12.5cm, angle=0]{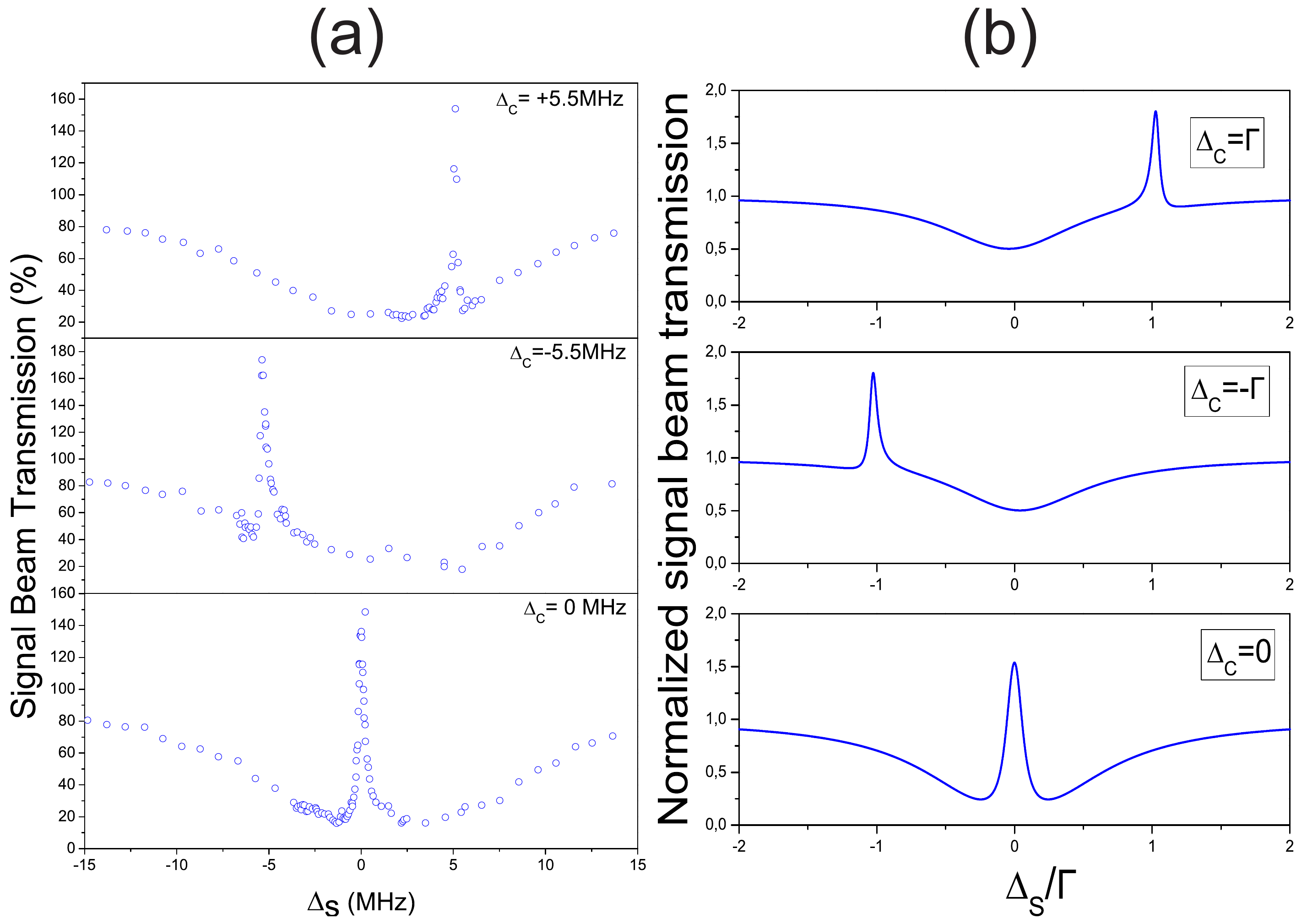}}
  \vspace{0.2cm}
  \caption{(a) Signal beam transmission as a function of the frequency detuning $\Delta_{S}$ for different values of the detuning $\Delta_{C}$. The beam $P$ is blue-detuned by $\Delta_P=1 MHz$ and the beams intensities are indicated in the text. (b) Normalized signal beam transmission spectrum calculated for three different detunings  of the coupling beam as indicated. The transmitted signal is normalized by the incident signal beam intensity. The used parameters, in units of $\Gamma$, are: $\Omega_C=0.1$, $\Gamma_P=0.05$}
    \label{fig:Fig2}
\end{figure}

\vspace{1.0cm}

	In a first set of measurements, we use all beams with gaussian modes and  investigate the Raman gain spectrum when the the two-photon detuning $\delta = \Delta_{C}-\Delta_{S}$ is scanned. Thus, for the pump beam $P$ blue-detuned by $\Delta_{P}= 1 MHz$, in Fig. 2(a) we show the signal transmission spectra for three different detunings of the coupling beam $C$, $i. e.,$ for $\Delta_{C}=0, \pm 5.5 MHz$. The spectra are obtained by changing the detuning  $\Delta_{S}$ in steps of $ 100 KHz$. As can be observed, a maximum gain of about $70\%$, for $\Delta_{C}= -5.5 MHz$, is observed around the two-photon resonance $\delta =0$. This gain is only observed in the presence of pump beam $P$ and depends on the intensity and detuning of this beam. On the other hand, its narrow  sub natural  linewidth indicates it is associated with the Zeeman ground-state coherence induced by the coupling and signal beams $C$ and $S$. For the spectra shown in Fig. 2(a), the corresponding intensities of beams $C$, $S$ and $P$ are $I_{C}=14 mW/cm^2, I_{S}=0.05 mW/cm^2$ and $I_{P}=10 mW/cm^2$, respectively. We understand this gain as originating from Raman amplification of the signal beam $S$ due to a population inversion between the two Zeeman ground states, induced by the pumping beam $P$. We have also checked  that although the pumping beam $P$ is necessary in order to observe the gain, it does not contribute to any parametric gain via FWM, since we choose its direction to avoid any phase-matching. In what follows, we model our system by considering the effect of the $P$ beam as an incoherent pump between consecutive ground state Zeeman sublevels.  

\subsection{Simplified theoretical model}

To theoretically model the main characteristics of the observed gain mechanism, we consider a homogeneously broadened three-level $\Lambda$ system as shown in Fig. 3. The states $\left| a \right\rangle$ and $\left| c \right\rangle$ belong to the non-relaxing ground states with zero energy. The state  $\left| b \right\rangle$ is the excited level with energy $\hbar \omega_{0}$ and decay rates to the two ground states equal to $\Gamma_{ba}$ and $\Gamma_{bc}$, with $\Gamma=\Gamma_{ba}+\Gamma_{bc}$ being the total spontaneous decay rate, $\Gamma/2\pi =5.2 MHz$. The incoherent pumping rate $\Gamma_{P}$ pumps directionally some of the population from state $\left| c \right\rangle$ to state $\left| a \right\rangle$. This incoherent pump rate is introduced to model the effect experimental beam $P$, which essentially transfers population from others Zeeman sublevels to the one interacting with the coupling beam C. This corresponds to a simplifying assumption, which, as we will see, captures the essencial physical characteristics of the observed results. A strong coupling field $C$ drives the transition $\left| a \right\rangle \longrightarrow \left| b \right\rangle$, while a weak signal field $S$ is applied to the transition $\left| c \right\rangle \longrightarrow \left| b \right\rangle$. The detunings of the $C$ and $S$ fields from the corresponding atomic resonant transitions  are denoted by $\Delta_{C}$ and $\Delta_{S}$, respectively.

\begin{figure}[htb]
  \centerline{\includegraphics[width=11.0cm]{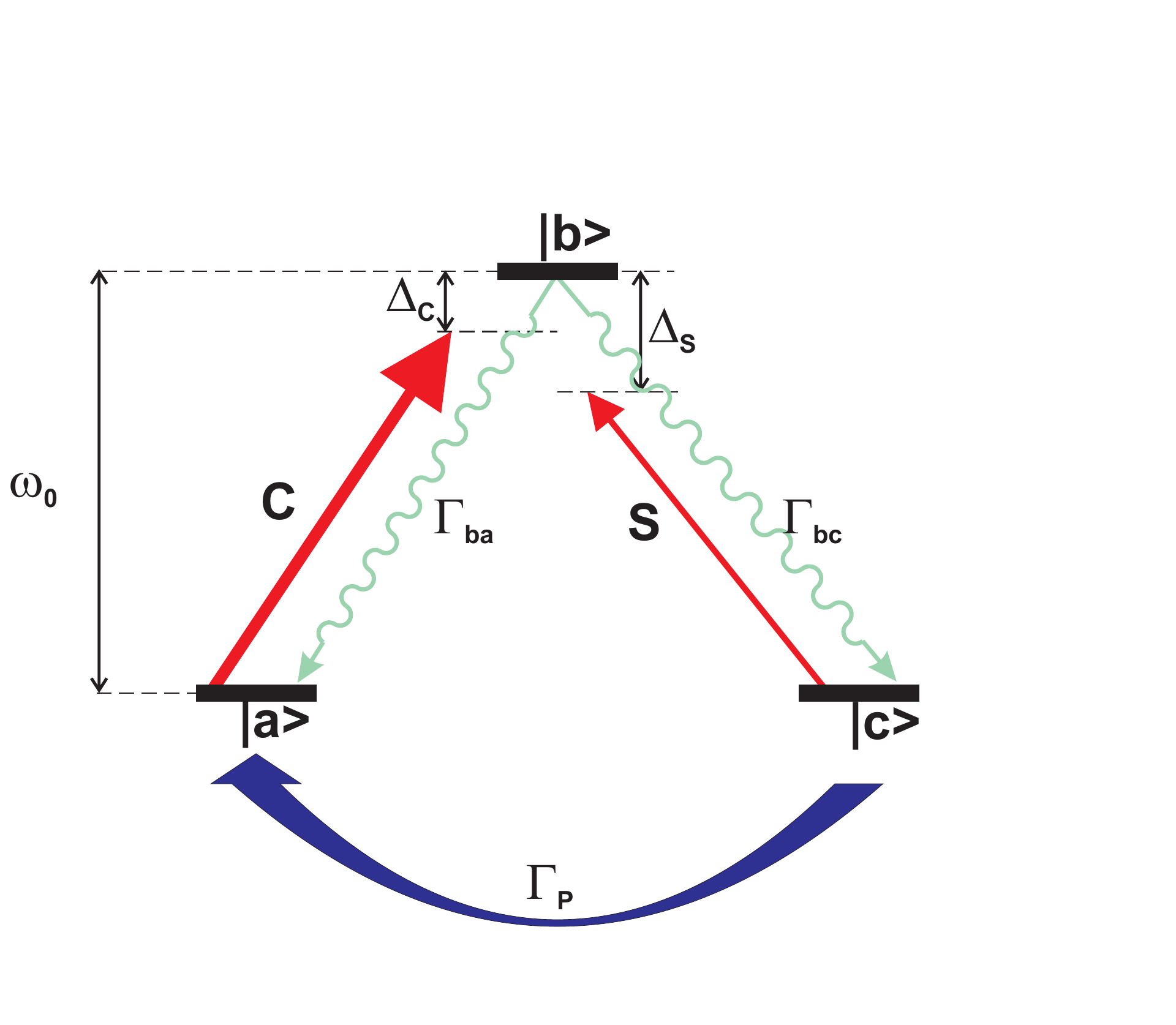}}
  \vspace{-0.5cm}
  \caption{Simplified $\Lambda$ three-level scheme interacting coherently with the beams $C$ and $S$. In the theoretical model we account for the beam $P$ through a directional incoherent pumping rate $\Gamma_{P}$ from state $|c\rangle$ to state $|a\rangle$. }
\label{fig:Fig1}
\end{figure}

In the rotating wave approximation, the time evolution for the
slowing varying population and coherence amplitudes ($\sigma$) are governed by the equations:

\begin{subequations}
\begin{align}
\dot{\sigma}_{aa} & = i[\Omega_{C} \sigma_{ba}-\Omega^{*}_{C}
\sigma_{ab}]+\Gamma_{ba}\sigma_{bb}+\Gamma_{P}\sigma_{cc},\\
\dot{\sigma}_{cc} & = i[\Omega_{S} \sigma_{bc}-\Omega^{*}_{S}
\sigma_{cb}]+\Gamma_{bc}\sigma_{bb}-\Gamma_{P}\sigma_{cc},\\
\dot{\sigma}_{bb} & = -i[\Omega_{C} \sigma_{ba}-\Omega^{*}_{C}
\sigma_{ab}+\Omega_{S} \sigma_{bc}-\Omega^{*}_{S}
\sigma_{cb}]-\Gamma\sigma_{bb},\\
\dot{\sigma}_{ab} & = i\Delta_{C}\sigma_{ab}+i\Omega_{C}(\sigma_{bb}-\sigma_{aa})-i\Omega_{S}\sigma_{ac}-\frac{\Gamma}{2}\sigma_{ab},\\
\dot{\sigma}_{cb} & = i\Delta_{C}\sigma_{cb}+i\Omega_{S}(\sigma_{bb}-\sigma_{cc})-i\Omega_{C}\sigma_{ca}-\frac{\Gamma}{2}\sigma_{cb},\\
\dot{\sigma}_{ac} & = i\delta\sigma_{ac}+i[\Omega_{C}\sigma_{bc}-\Omega^{*}_{S}\sigma_{ab}]-\frac{\Gamma_{P}}{2}\sigma_{ac}.
\end{align}
\end{subequations}

In the above equations we consider $\Gamma_{ba}=\Gamma_{bc}=\Gamma/2$, and $\delta = \Delta_{C}-\Delta_{S}$.  $\Omega_i$$ (i=C, S)$
represents the Rabi frequency associated with beam $i$ in an atom
at position $\vec{r}$, and is given by
$\Omega_i(\vec{r})=\mu_{i} {\cal E}_i(\vec{r})e^{i
\vec{k}_i\cdot \vec{r}}/\hbar$, with $\mu_{i}$ being the dipole moment
of the transition coupled by $\Omega_i$, ${\cal
E}_i(\vec{r})$ the electric field envelope of beam $i$, and
$\vec{k}_i$ its wave-vector. 

	We assume a closed three-level system, so that $\sigma_{aa}+\sigma_{bb}+\sigma_{cc}=1$. In order to find the response of the system, we solve the above equations, in steady state regime, in all orders in the coupling field $C$ and in first-order in the signal field $S$, to obtain $\sigma^{(1)}_{ba}$ in terms of the zero-order solutions $\sigma^{(0)}_{ij}$:

\begin{equation}
\sigma^{(1)}_{cb}=\frac{\Omega_{S}[-\delta+i\frac{\Gamma_{P}}{2}](\sigma^{(0)}_{cc}-\sigma^{(0)}_{bb})-\Omega_{C}\Omega_{S}\sigma^{(0)}_{ba}}{[i\delta+\frac{\Gamma_{P}}{2}](i\Delta_{S}-\frac{\Gamma}{2})-\Omega_{C}^{2}}
\end{equation}
 
The relevant zero-order populations and coherence are given by:
\begin{subequations} \label{order0}
\begin{align}
\sigma^{(0)}_{aa} & = \frac{\Omega_{C}^{2}+\Delta_{C}^{2}+\Gamma^{2}/4}{\Omega_{C}^{2}(2+\Gamma /2 \Gamma_{P})+\Delta_{C}^{2}+\Gamma^{2}/4}\label{order0a},\\
\sigma^{(0)}_{bb} & =\frac{1-\sigma^{(0)}_{aa}}{1+\Gamma /2 \Gamma_{P}},\\
\sigma^{(0)}_{cc} & =\frac{\Gamma}{2\Gamma_{P}}\frac{1-\sigma^{(0)}_{aa}}{1+\Gamma /2 \Gamma_{P}},\\
\sigma^{(0)}_{ab} & =-i\Omega_{C}\frac{1-(2+\Gamma /2 \Gamma_{P})\sigma^{(0)}_{aa}}{(i\Delta_{C}-\Gamma/2)(1+\Gamma /2 \Gamma_{P})}.
\end{align}
\end{subequations}

	The absorption of the signal beam is determined by the imaginary part of $\sigma^{(1)}_{cb}$. In Fig. 2(b) we plot the signal beam transmission spectrum, in units of $\Gamma$, for three different coupling beam detunings, as used in the experiment. For these spectra we have set the optical density of the medium equal to 3, which is of the order of the one measured experimentally, and used $\Omega_C/\Gamma =0.1$ and $\Gamma_P/\Gamma =0.05$. As we can see, this simple model reproduces well the main characteristics of the measured spectra. We have checked that for $\Gamma_P=0$, the corresponding spectra reduce to the usual $EIT$ spectra, showing total transparency at $\delta=0$, but no gain. Nevertheless, in order to reproduce the observed spectrum the used Rabi frequency for the coupling beam is about one order of magnitude smaller than the one used in the experiment. This discrepancy can be due to the simplification of our atomic level scheme, which does not accounts for the total multiplicity of Zeeman sublevels, and which can lead to a much smaller effective intensity. Moreover, the optical coherence $\sigma^{(1)}_{cb}$ is proportional to the complex Rabi frequency associated with beam $S$, therefore, it has the same phase dependence of this incident optical field. We have also measured the dependence on the gain with the intensity of the coupling beam, and the theoretical model also predicts qualitatively  the observed saturation behavior.

\subsection{Amplification of $LG$ mode}

Now we turn to the observation of narrow band amplification of a LG mode carrying OAM. For this purpose, we use a spatial light modulator ($SLM$), not shown in the experimental scheme of Fig. 1, to shape the signal beam $S$ in a $LG$ mode with topological charge $\ell$. For such mode the electric field amplitude, described in the cylindrical coordinates $(\rho, \phi, z)$ at the plane $z=0$, is given by:

\begin{equation}
{\cal E}_{S}(\vec{r})=LG_{p}^{\ell}(\rho,\phi)={\cal E}_0\left(\frac{\rho \sqrt2}{ w_0}\right)^{|\ell|}e^{-\frac{\rho^2}{ w_0}}e^{i\ell\phi}L_{p}^{|\ell|}\left(\frac{2\rho^2}{w_0^{2}}\right),
\end{equation}
where $w_0$ and ${\cal E}_0$ are, respectively, the beam waist and amplitude,  and $L_{p}^{|\ell|}$ the associated Laguerre polynomial with radial index $p$. According to our experimental situation, we consider only single annular LG modes with $p=0$ for the $S$ beam, with the incident coupling beam $C$ and the pump beam $P$ as gaussian modes with waists larger than that of the $S$ beam.  As we demonstrated in the previous subsection, the coherence responsible for the amplification of the signal beam $S$ is proportional to the complex amplitude of this mode and therefore should retain the spatial phase structure of the mode.\\
\begin{figure}[htb]
  \centerline{\includegraphics[width=9.0cm, angle=0]{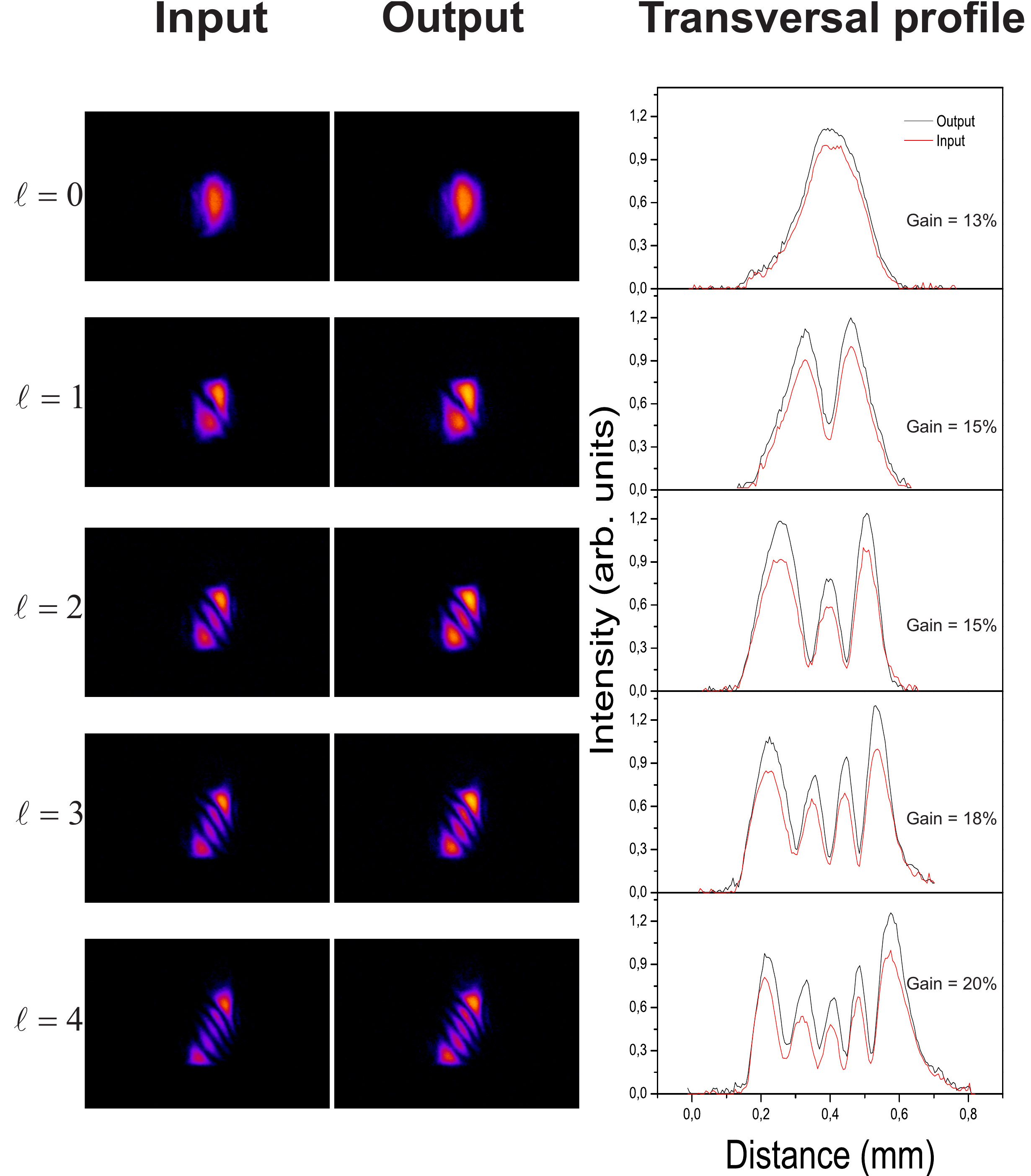}}
  \vspace{-0.2cm}
  \caption{Measurement of the topological charge of the signal beam $S$. First column: Topological charge of the incident beam, measured without the atoms. Second column: Topological charge of the amplified signal beam via the Raman gain mechanism. Third column: Transversal profile of the images shown in the first column (red) and in the second column (black). For these data, the beams $C$ and $S$ are on Raman resonance and red-detuned by 6.8 MHz, while the $P$ beam is on resonance.}
  \label{fig:Fig4}
\end{figure}
	
	To verify this fact experimentally, as indicated in Fig. 1, we also recorded the signal beam transmission in a CCD camera, with and without the MOT. In order to optimize the interaction region in the atomic cloud and also the $LG$ beam mode, we have adjusted the waist of the $LG$ mode such that it increases slightly between consecutive orders. The measured beam waists at the atomic cloud are 0.3 mm, 0.5 mm, 0.7 mm 1.0 mm and 1.1 mm, respectively for the modes with $\ell=0, 1, 2, 3, 4$. All these values are much smaller than the waists of  the $C$ and $P$ beams, as well as the size of the atomic cloud. This explains the slight increase in the observed gain with the order of the $LG$ beam, as indicated in the third column of Fig 4. The measured gains are $13\%$, $15\%$, $15\%$,  $18\%$ and $20\%$, respectively for $\ell=0, 1, 2, 3, 4$.   

	In order to demonstrate that the amplified transmitted beam is still in the same $LG$ mode, we used the tilted lens method described in \cite{Vasnetsov04, Singh13}. According to this method, the self-interference pattern of a LG beam with topological charge $\ell$, observed after the astigmatic optics, exhibits $|\ell|+1$  bright fringes turned by an angle whose sign is the sign of $\ell$. In the first column of Fig. 4 we show the corresponding images for the five different values of the topological charge for the case where the MOT is switched off. These images then reveal the topological charge of the incident signal beam $S$. When the atoms are present, we show in the second column of Fig. 4 the corresponding recorded  amplified images, which preserve the same value of the topological charge. In order to analyze quantitatively these observations we use as  figure of merit the contrast associated with each peak of the images. Thus, we have plotted in the third column of Fig. 4 the transversal profile of the transmitted beams in both cases and verified that the peak contrast is approximately the same for all peaks in the original and amplified images. Finally, it is worth to mention that we have also experimentally verified that when beam $C$ is the one carrying OAM, this OAM is not transferred to the signal beam $S$.

\section{Conclusions}

\vspace{-0.0cm}
We have experimentally observed a stimulated narrow-band Raman gain mechanism leading to the amplification of a weak signal beam in an ensemble of cold cesium atoms. This gain mechanism was modeled using a simple $\Lambda$ three-level system interacting with two coherent fields in the presence of incoherent pumping. The theoretical model reproduces qualitatively well the main characteristics of the observed Raman gain spectrum, as, for example, its amplitude and subnatural spectral width. 
 We have used this gain mechanism to demonstrate, for the first time, the narrow band amplification of a $LG$ mode carrying OAM. These results are of considerable importance owing to the recent demonstration of efficient classical communication using light modes with OAM-encoded information. Long distance communication using light carrying OAM will certainly require amplification nodes, or repeaters, which preserves the phase structure of the mode, to compensate for inevitable losses. Moreover, the demonstrated narrow band, subnatural, gain mechanism constitute by itself an alternative way to filter out signal from noise in such communication systems. \\

\section*{Acknowledgment}
We acknowledge R. A. de Oliveira and W. S. Martins for experimental assistance in the early stage of this experiment and A. Z. Khoury for stimulating discussions.
This work was supported by the Brazilian agencies CNPq and FACEPE. We also thank CAPES-COFECUB (Ph 740-12) for the support of Brazil-France cooperation.
                                                                                                    
\end{document}